\documentclass{ws-ijmpa}
\usepackage[super,compress]{cite}
\usepackage{graphicx}
\begin{document}
\markboth{ D.P. Kirilova, E.M. Chizhov }{Cosmological Constraints on
Chiral Tensor Particles}

%
%

\title{Cosmological Constraints on Chiral Tensor Particles}

\author{D. P. Kirilova}

\address{Institute of Astronomy and National Astronomical Observatory, Bulgarian Academy of Sciences,
1784 Sofia, Bulgaria, dani@astro.bas.bg}

\author{E. M. Chizhov}

\address{Physical Faculty, Sofia University,
1164 Sofia, Bulgaria}

\maketitle

\begin{history}
\received{Day Month Year} \revised{Day Month Year}
\end{history}

\begin{abstract}

 We discuss an extended
model with chiral tensor particles in the Universe. Their direct
influence on the Universe dynamics and their characteristic
interactions in the hot Universe plasma, considered in previous
publications, are briefly reviewed. A short discussion on the
contemporary cosmological bounds on effective number of the
relativistic degrees of freedom is provided. Cosmological
constraints on the tensor particles interactions strength are
obtained, corresponding to different cosmological bounds on the
relativistic degrees of freedom and for different assumptions about
right handed neutrinos.

\keywords{chiral tensor particles; early Universe; BBN constraint}

\end{abstract}
\ccode{PACS numbers: 98, 14}

\section{Introduction}

The extended model with chiral tensor (ChT) particles, new type of
spin-1 particles, was first discussed by Chizhov~\cite{MPLA}. The
ChT particles were introduced as an extension of the Standard Model
for completeness of the representation of the Lorentz group. These
particles are predicted to be the carriers of new interaction and
they have only chiral interactions with the known fermions, through
tensor anomalous coupling. This model and its predictions, as well
as the signatures for the experimental detection of ChT particles
have been discussed in numerous
publications~\cite{EChAYa,EChAYaa,LHC,LHCb,mih2009}.
Experimental constraints on ChT particles masses and couplings
provided by the ATLAS collaboration at the Large Hadron Collider at
CERN, that carries on an experimental search of ChT particles.

 The cosmological influence
of ChT particles has been analyzed, as well
(see refs.~\cite{todor,AATr,IJMPA,MPLA17}
In these works ChT particles influence on the
Universe dynamics and their characteristic interactions with the
early Universe plasma were considered.

The ChT particles and the corresponding extension of the Higgs
sector contribute to the matter tensor in the right-hand side of the
Einstein--Hilbert equation, increasing the Universe density and
changing the dynamical evolution of the Universe. Namely, at the
early epoch while ChT particles are relativistic, their additional
degrees of freedom and the ones introduced by the corresponding
Higgs sector extension in the model increase the total effective
degrees of freedom by $g_{ChT}=28$, corresponding to two additional
tensor doublets, a triplet and singlet gauge vector particles and an
extra Higgs doublet. This results into a slight increase of the
expansion rate during the period of effective interactions of ChT
particles in comparison with the Standard Cosmological Model.

The characteristic interactions of ChT particles with the early
Universe plasma, their creation, scattering, annihilation and decay
processes were updated recently~\cite{MPLA17} using the latest
experimental constraints on their characteristics, that have been
obtained at ATLAS (Aad et al., 2014, 2014a)~\cite{ATLAS14,ATLAS14a}
at LHC. The constraints on the tensor particle masses, provided by
the ATLAS Collaboration read $M_T^0>2.85$ TeV and $M_T^+>3.21$ TeV.

The time interval of effective interactions of ChT particles in the
early Universe history and, hence, of their abundant presence was
determined to be~\cite{MPLA17}:
$$t_c=6\times 10^{-42}~{\rm s}<t<t_d=5\times
10^{-14}~{\rm s}.$$
 The corresponding energy range is from
$T_c=1.8\times 10^{17}$~GeV down to $T_d=6 \times 10^3$~GeV. ChT
creation processes unfreeze at $t>t_c$, when the Universe
temperature falls below $T_c$, where $T_c$ has been estimated by
comparing the tensor particle creation rate with the universe
expansion rate at the radiation dominated (RD) epoch $H=\sqrt{8\pi^3
G_N g_*/90}\:T^2$, where $g_*$ is the total number of the effective
degrees of freedom~\cite{MPLA17}. In most inflationary scenarios,
however, the temperature at the reheating epoch after inflation is
much lower. Thus, actually tensor particles creation processes
proceed effectively since the reheating epoch of the concrete
inflationary model till ChT decay at around $T_d=6\times 10^3$~GeV.

 Chiral tensor particles are present during a period much earlier than
 the Big Bang Nucleosynthesis
(BBN) epoch and Cosmic Microwave Background (CMB) formation epoch,
and, therefore, ChT particles cannot directly influence the
observable relics from BBN and CMB - the light element abundances
and the observed characteristics of CMB.

However, in case ChT particles interact with the right-handed
neutrinos (or other representatives of dark radiation){\it indirect
constraints} on ChT interactions can be obtained from the
requirement that the introduced into equilibrium additional
relativistic right-handed neutrinos should not exceed the
cosmological (BBN, CMB, etc.) limits.

In the next section we discuss different recent cosmological
constraints on the effective number of relativistic particles. In
the third section we discuss the possible ChT particles interactions
with right-handed neutrinos and obtain a cosmological constraint on
the ChT particle interactions strength for different number of light
right handed neutrino species and for different cosmological bounds
on the effective number of relativistic particle species during BBN.

\section{BBN and cosmological constraints on additional radiation}

 Big Bang Nucleosynthesis is a reliable probe of the physical
 conditions of the early universe and provides a unique test of
 Physics Beyond the SM during the radiation dominated epoch.
 See for example refs.~\cite{Cyburt,Foley,Coc,BBN}.

 Contemporary BBN is theoretically well established - based on
 well-understood physics. The primordially produced abundances are
 usually parameterized by
the baryon-to-photon ratio, relativistic energy density (effective
number of relativistic
neutrino types $N_{eff}$) and neutron life time. 
The predicted abundances depend also on the well measured cross
sections of nuclear processes, which have been continuously
 updated~\cite{Angulo,Xu}. 
More and more precise BBN codes have been
invented~\cite{Pisanti,Consiglio,Arbey,Pitrou}.

Precision Planck CMB measurements also probe the baryon density,
helium content, and $N_{eff}$, thus allowing for an independent
precise cosmological test using CMB data alone ~\cite{Cyburt}.

The precision of observational data on primordial abundances has
been drastically improved, as well, during the last decade. New
observations of high redshif low metalicity QSO help to improve
considerably the determination of D and its
uncertainty~\cite{Cooke}. 
During recent years more precise determination of He-4 abundance was
provided, as well, thanks to its updated emissivity and the
observed new infra-red line ~\cite{Izotov}.

BBN predicted abundances (except Li-7) are in a good overall
agreement with the ones inferred from observations. This good
concordance between BBN theory and observational data allows to use
it as a precision probe of Beyond SM Physics during the BBN epoch (1
MeV- 10 KeV).

\subsection{Cosmological constraints on the effective number of
relativistic particles}

 In particular, BBN is used as a precise
speedometer at RD stage. He-4 is very sensitive element to the
expansion rate of the universe at the BBN epoch $H(t)$, which is
usually parameterized by the effective number of relativistic
neutrino types $N_{eff}$ (Shvartsman 1969)~\cite{Shvartsman}. At RD
stage of the universe relativistic neutrinos (or any additional
relativistic particles) contribute to the total energy density
$\rho$ by:
\begin{equation}
\rho_{\nu}=7/8(T_{\nu}/T)^4N_{eff}\rho_{\gamma}(T).
\end{equation}
Hence, they influence the expansion rate of the universe $H \sim
\sqrt{8 \pi G_N \rho/3}$, and the primordial production of light
elements.

 Hence, the cosmological bounds on
$\delta N_{eff}$ are used to constrain Beyond SM physics which
introduce additional light species, as for example supersymmetric
scenarios (constraining lightest particle neutralino or gravitino),
string theory, extra dimensions theories, theories with right-handed
(sterile) neutrinos, neutrino oscillations, lepton asymmetry, the
discussed here chiral tensor particles, etc.

We list below some of the recently obtained cosmological bounds on
$N_{eff}$.

 A maximum likelihood analysis by Cyburt et al. (2016)~\cite{Cyburt},
 provides the following BBN constraints: $N_{eff}<3.4$.

   Recently He-4 primordial mass fraction was determined with even
   higher accuracy (Pitrou et al. (2018))~\cite{Pitrou}:

  $$Y_p=0,24709 \pm 0,00017.$$

This allowed to put the following stringent constraints on $N_{eff}$
by cosmological considerations of the BBN produced He-4:

   $$N_{eff}=2.88 \pm 0.27~~(95\%).$$

For comparison the CMB constraint (Planck Collaboration
2015)~\cite{Ade} reads:

                                  $$N_{eff}=3.13 \pm 0.31~~(95\%).$$

Besides, constraints on $N_{eff}$ have also been obtained on the
basis of Lyman Alpha forest BOSS data, CMB data from Planck, ACT,
SPT, WMAP polarization (Rossi Yeche et al. 2015)~\cite{Rossi}:
  $$N_{eff}=2.911 \pm 0.22~~(95\%).$$
for neutrino mass $m_{\nu} <0.15$ eV.  Similar stringent constraints
 from baryon acoustic oscillations data are discussed by Sasankan et al.
 2017~\cite{Sasankan}.

The improved determination of D provides even more stringent
constraints. D/H provides a tight measurement of $N_{eff}$ when
combined with the CMB baryon density and provides a $2\sigma$ upper
limit $N_{eff}<3.2$ ~\cite{Cyburt}.

Using CMB plus D plus He-4 data allows a considerable reduction of
the error~\cite{Cyburt}: 

 $$N_{eff}=2.88 \pm 0.16~~(95\%~~ \rm{Planck+D+He-4})$$

Similar stringent constraint is provided by CMB plus BBN analysis:

                  $$N_{eff}=3.01 \pm 0.15~~(95\% ~~\rm{Planck+BBN})$$

  In particular, cosmology constraints
severely the thermalized during BBN light sterile neutrinos. In the
next section these cosmological limits are used to constrain the
decoupling temperature of right handed neutrinos and ChT particles
interaction strength.

\section{BBN Constraints on the ChT Particles Interactions Strength}

If right-handed neutrinos interact with the chiral tensor particles
they may be produced through ChT particles exchange during BBN
epoch. The term of the effective Lagrangian corresponding to the
right handed neutrino coupling reads:

$$L=\frac{4}{\sqrt{2}} G_T \bar{e}_L\sigma_{\alpha \beta} \nu_R ~. ~\bar{u}_L \sigma_{\alpha \beta} d_R +h.c.,$$
where $G_T$ measures the effective ChT interaction strength,
$\sigma_{\alpha \beta}=\frac{i}{2}(\gamma_{\alpha}\gamma_{\beta} -
\gamma_{\beta} \gamma_{\alpha}).$

It is straightforward to obtain cosmological bound on the coupling
constant of new particles on the basis of BBN
considerations~\cite{dolgov}. Cosmological constraint on $G_T$ was
first discussed in
ref.~\cite{blois}. 

In what follows we obtain cosmological constraints on ChT
interaction strength $G_T$ for different number of light right
handed neutrinos and using different cosmological constraints on the
effective number of relativistic particle species $\delta N_{eff}$.
Assuming that ChT particles interact with right handed neutrinos, we
determine also the decoupling temperature of right-handed neutrinos
in different cases.

\subsection
{\bf $\delta N_{eff}<0.4.$}

  At present the most conservative BBN bound on the additional relativistic (light) neutrinos is:
\begin{equation}
\delta N_{eff}=g_R(T_{\nu_R}/T_{\nu_L})^4<0.4.
\end{equation}

Then in case of three light right-handed neutrinos $\nu_R$ , it
follows
\begin{equation}
3(T_{\nu_R}/T_{\nu_L})^4<0.4,
\end{equation}
which puts a constraint on the decoupling/freezing temperature of
the right-handed neutrinos $T_f$ using the BBN constraint on $\delta
N_{eff}$ and entropy conservation relation $g T^3=const$, namely
\begin{equation}
T_{\nu_R}/T_{\nu_L}=(\frac{43}{4}/g_*(T_f))^{1/3}<0.604.
\end{equation}
The following constraint on the decoupling temperature is obtained:
\begin{equation}
T_f>251~\rm{MeV}.
\end{equation}
The decoupling temperature of a given species is connected with its
interactions coupling strength:
\begin{equation}
(G_T/G_F)^2\sim(T_f/2~{\rm MeV})^{-3},
\end{equation}
 where $2$ MeV is
the decoupling temperature of the active neutrino species, $G_F$ is
the Fermi constant. Hence, the constraint on the ChT particles
coupling is derived:
\begin{equation}
G_T<7.1\times 10^{-4}G_F.
\end{equation}

In case of 2 light sterile neutrinos the decoupling temperature is
$T_f>177~\rm{MeV}$ and the corresponding constraint on ChT particles
coupling reads: $G_T<1.19\times 10^{-3}G_F$. For only 1 light
sterile neutrino the sonstraints are, correspondingly:
$T_f>125~\rm{MeV}$ and $G_T<2.0\times 10^{-3}G_F$.

\subsection
{\bf $\delta N_{eff}<0.3.$}

Using the bound:
\begin{equation}
\delta N_{eff}=g_R(T_{\nu_R}/T_{\nu_L})^4<0.3,
\end{equation}
which is close to the recently obtained refined BBN bound on the
additional light neutrinos~\cite{Pitrou}, we determined the
cosmological constraint on ChT  interaction strength in case of 3, 2
and 1 light sterile neutrino. Following the described above
considerations, in case of tree light $\nu_R$ the decoupling
temperature of $\nu_R$ is found to be $T_f>354~\rm{MeV}$ and the BBN
constraint on ChT particles strength is
\begin{equation}
 G_T<4.2\times 10^{-4}G_F.
\end{equation}

In case of 2 light $\nu_R$, we have found $T_f>224~\rm{MeV}$ and
$G_T<8.4\times 10^{-4}G_F$, correspondingly.

In case of 1 light $\nu_R$, $T_f>158~\rm{MeV}$ and
  $G_T<1.4 \times 10^{-3}G_F$.

\subsection
{\bf $\delta N_{eff}<0.2.$}

 For the more stringent BBN constraint
\begin{equation}
\delta N_{eff}=g_R(T_{\nu_R}/T_{\nu_L})^4<0.2,
\end{equation}
obtained when D data is also included~\cite{Cyburt}, the
cosmological constraints on $G_T$ for 3, 2 and 1 $\nu_R$ are
correspondingly:

In case of tree light $\nu_R$, $T_f>1585~\rm{MeV}$
\begin{equation}
  G_T<4.5 \times 10^{-5}G_F.
\end{equation}

In case of 2 light $\nu_R$, $T_f>354~\rm{MeV}$ and $G_T<4.2 \times
10^{-4}G_F$.

In case of one light $\nu_R$, $T_f>178~\rm{MeV}$ and $G_T<1.2 \times
10^{-3}G_F$.

Thus, contemporary precise cosmological data, and in particular the
improved bounds on $\delta N_{eff}$, leads to more stringent
constraints on ChT particles interactions - they should be
milli-weak or weaker (depending on the assumed number of light
right-handed neutrino species). Therefore, for all analyzed cases
the derived cosmological  constraints require tensor particles
masses higher than several TeV, while in many cases they should be
higher than tens of TeV. Thus, the obtained here cosmological
constraints are stronger than the experimental constraints on the
tensor particle masses, provided by the ATLAS Collaboration at LHC.

\section{Conclusion and Discussion}

We discuss an extended Beyond Standard Model of Particle Physics and
Cosmology with new ChT particles. The presence of these particles
during the very early universe leads to slight speed up of the
universe expansion and ChT particles direct interactions in the
early universe plasma: creation, scattering, annihilation and decay.
However, it was found that ChT particles have been abundant at such
an early stage of universe evolution, that they or their decay
products have not effected BBN, CMB or other processes which have
left observable relics in today's universe. Thus, ChT particles are
allowed from
     cosmological considerations, however, direct cosmological constraints cannot
     be obtained.

However, indirect cosmological constraints on ChT interactions can
 be derived in case ChT particles interact with right handed neutrinos.
In this work we discuss this possibility. We have derived stringent
cosmological constraints on ChT interactions strength - it should be
milli weak or weaker, depending on the number of the light
right-handed neutrino types. For all analyzed cases tensor particles
masses should be higher than several TeV, while for many cases they
should be higher than tens of TeV. Thus, the obtained here
cosmological constraints are in accord or stronger than the
experimental constraints on the tensor particle masses, provided by
the ATLAS Collaboration at LHC. The eventual experimental detection
of the chiral tensor particles will be of great interest because it
will reveal the realm of milli weak interactions.

These cosmological constraints can be interpreted also as an
indication for absence of ChT interactions with sterile
neutrinos. 
On the contrary, eventual future detection of the ChT particles and
determination of their interaction strength may be used as an
indicator for the number of the light right-handed neutrino species.

\section*{Acknowledgments}
We thank M.Chizhov for useful discussions concerning ChT model and
ChT phenomenology. We acknowledge the partial financial support by
projects DN 08-1/2016 and DN18/13-12.12.2017 of the Bulgarian
National Science Fund of the Bulgarian Ministry of Education and
Science. We appreciate the useful suggestions of the unknown
referee, which helped to improve the quality of this work.

\end{document}